**Opportunities for Mining Radiology Archives for Pediatric Control Images**

Camilo Bermudez[1*], Varvara N Probst[2], Larry T Davis[3], Thomas Lasko[4], Bennett A Landman[1,2]

[2] Electrical Engineering, Vanderbilt University, Nashville, TN

[1] Biomedical Engineering, Vanderbilt University, Nashville, TN

[3] Department of Radiology, Vanderbilt University, Nashville, TN

[4] Department of Biomedical Informatics, Vanderbilt University, Nashville, TN

* Corresponding author

**Abstract**

A large database of brain imaging data from healthy, normal controls is useful to describe physiologic and pathologic structural changes at a population scale. In particular, these data can provide information about structural changes throughout development and aging. However, scarcity of control data as well as technical challenges during imaging acquisition has made it difficult to collect large amounts of data in a healthy pediatric population. In this study, we search the medical record at Vanderbilt University Medical Center for pediatric patients who received brain imaging, either CT or MRI, according to 7 common complaints: headache, seizure, altered level of consciousness, nausea and vomiting, dizziness, head injury, and gait abnormalities in order to find the percent of studies that demonstrated pathologic findings. Using a text-search based algorithm, we show that an average of 59.3% of MRI studies and 37.3% of CT scans are classified as normal, resulting in the production of thousands of normal images. These results suggest there is a wealth of pediatric imaging control data which can be used to create normative descriptions of development as well as to establish biomarkers for disease.

**Introduction**

Understanding how healthy brains change over time provides insight into development, aging, and disease. However, recent publications have pointed out a deficiency of imaging data for healthy, normal controls in children, as well as practical and technical challenges for imaging in the young [1-3]. A large database with cross-sectional samples of control populations is essential for the characterization of trajectories in development and aging. These claims of the inadequacy of controls led us to explore the medical record to show that there are, in fact, suitable numbers of imaging controls for any given complaint. We propose a text-based, complaint-specific method of searching the medical record for negative studies. Our results show that negative findings on imaging are common enough to be used as controls for common clinical complaints such as headache, seizure, altered level of consciousness, dizziness and giddiness, nausea and vomiting, and abnormality of gait.

**Methods**

In order to find the number of normal and abnormal imaging studies, we used internally developed software that searches the Vanderbilt University Medical Center electronic medical record (EMR) by demographics, billing codes (ICD-9, ICD-10, PheWAS, CPT) and clinical data (vital signs, laboratory results, medications). First, we filtered the EMR by age to select for patients under 18 years. Next, we filtered those results by patients who received either head CT or brain MRI. We then filtered by complaints and symptoms that would likely prompt clinicians to seek imaging in pediatric patients - headache, seizures, alteration of consciousness, vomiting and nausea, dizziness and giddiness, concussion and post-concussion syndrome and abnormality of gait. These complaints would benefit from the development of developmental trajectories to better understand their involvement in pathologic processes of the young.

To sort out the healthy images from this set, we performed a text search through the radiology reports with phrases associated with normal radiology reports such as "Normal study," "Normal Brain MRI," "Normal head CT," "No findings of intracranial pathology," and others (Appendix A, B). This over-sensitive approach was intended to include any report that may be normal. The total number of healthy imaging studies was compared to the number of abnormal imaging studies (Figure 1).

In order to validate whether this method would accurately yield normal studies, the subset of "healthy controls" were manually inspected by reading the anonymized radiology report for the respective examination. One hundred reports were randomly selected for inspection and identified as normal or abnormal. These records were obtained through the Synthetic Derivative, a de-identified database of EMR under approval of the Vanderbilt University Institutional Review Board. The 95% confidence intervals for the number of normal reports were estimated assuming a binomial distribution of normal and abnormal radiology reports.

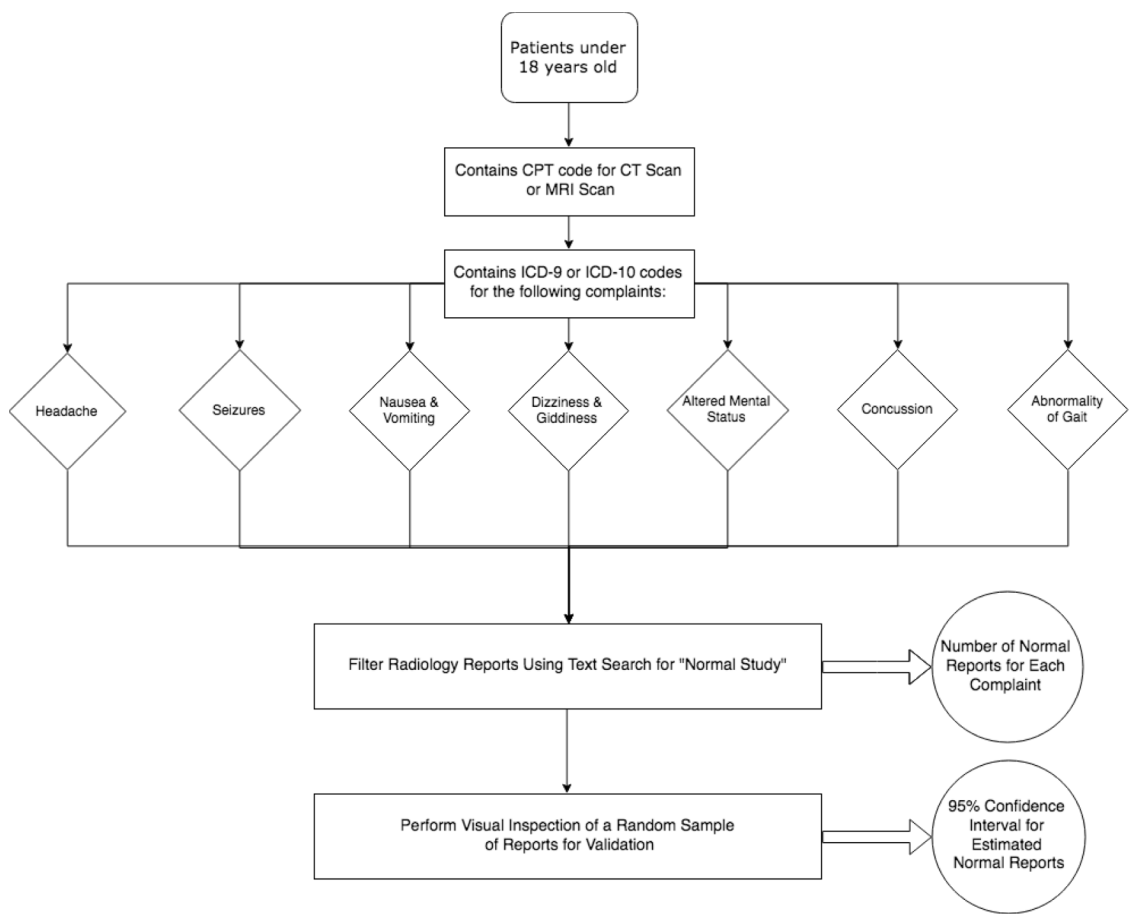

**Figure 1:** Flow chart of filtering process to estimate the number of normal radiologic studies under 7 common clinical complaints.

**Results**

Mining of the EMR showed a large number of pediatric patients that received imaging without any significant pathologic findings. Our method estimates that, on average, 59.3% of brain MRI studies obtained for the seven chosen complaints were interpreted as normal by the radiologist, yielding thousands of normal MRI studies in subjects under 18 years old (Table 1). Validation shows that the average rate is within 35.2% to 46.4%, resulting in hundreds, if not thousands, of normal studies for each complaint. Similarly, the estimated average rate of normal head CT studies using our method was 37.3%. Validation shows an average 95% confidence interval of 28.2% to 33.6%, resulting in hundreds or thousands of normal studies.

**Discussion & Conclusions**

There is a wealth of control data produced by normal radiological exams obtained in pediatric patients. While the rate of normal imaging varies by clinical complaint, manual inspection showed that there is still a large amount of normal radiologic data in a pediatric population. Our search included some of the most common complaints but was by no means an exhaustive exploration of clinical indications for imaging or abnormal findings. Further work for quality assurance or conservative restrictions on data selection would be required, particularly for CT, which showed a lower rate of correct selection. However, the benefit of a large control sample is the opportunity to build a pediatric brain growth atlas that can describe normative development of the brain. Developing a large database of multimodal MRI sequences and CT scans in a demographically diverse sample will enable the detection of subtle and important differences in brain developmental trajectories in individuals with various neurological conditions [4]. Furthermore, we will be able to establish imaging biomarkers in order to predict outcomes and improve prognosis in children with various neurological conditions [4].

**Table 1:** Number of total and normal MRI studies in pediatric patients with common clinical complaints.

| Symptoms | Number of Subjects with Complaint | Number of Subjects with Complaint that Received MRI | Number of MRI Read as Normal | 95% Confidence Interval after Validation |
|---|---|---|---|---|
| Headache | 18,396 | 4,604 (25.2%) | 3,182 (69.1%) | [2218, 2753] (48.2% - 59.8%) |
| Seizures | 9917 | 5478 (55.2%) | 3420 (62.4%) | [2053, 2694] (37.5% - 49.2%) |
| Altered Mental Status | 4,779 | 2,001 (41.9%) | 1,288 (64.4%) | [856, 1082] (42.8% - 54.1%) |
| Vomiting & Nausea | 55,905 | 4998 (8.94%) | 3140 (62.8%) | [1754, 2361] (35.1% - 47.2%) |
| Dizziness & Giddiness | 2316 | 528 (22.8%) | 129 (24.4%) | [90, 112] (17.0% - 21.1%) |
| Concussion | 4211 | 405 (9.62%) | 273 (67.4%) | [175, 225] (43.2% - 55.6%) |
| Abnormality of Gait | 5112 | 1610 (31.5%) | 1044 (64.9%) | [396, 608] (22.9% - 37.8%) |

**Table 2:** Number of total and normal CT studies in pediatric patients with common clinical complaints.

| Symptoms | Number of Subjects with Complaint | Number of Subjects with Complaint that Received CT | Number of CT Read as Normal | 95% Confidence Interval after Validation |
|---|---|---|---|---|
| Headache | 18,396 | 4401 (23.9%) | 3525 (80.1%) | [2419, 3020] (55.0% - 68.6%) |
| Seizures | 9,917 | 3087 (31.1%) | 1000 (32.4%) | [697, 865] (22.6% - 28.0%) |
| Altered Mental Status | 4,779 | 2800 (58.6%) | 611 (21.8%) | [503, 581] (18.0% - 20.8%) |
| Vomiting & Nausea | 55905 | 7172 (12.8%) | 1395 (19.5%) | [1166, 1336] (15.6% - 18.6%) |
| Dizziness & Giddiness | 2316 | 442 (19.1%) | 215 (48.6%) | [180, 206] (40.7% - 46.6%) |
| Concussion | 4211 | 2262 (53.7%) | 530 (23.3%) | [477, 497] (21.1% - 22.0%) |
| Abnormality of Gait | 5112 | 1016 (20.0%) | 360 (35.4%) | [251, 311] (24.7% - 30.6%) |

**Acknowledgement**

This research was supported by NSF CAREER 1452485 (Landman). This study was also supported in part using the resources of the Advanced Computing Center for Research and Education (ACCRE) at Vanderbilt University, Nashville, TN. This project was supported in part by the National Center for Research Resources, Grant UL1 RR024975-01, and is now at the National Center for Advancing Translational Sciences, Grant 2 UL1 TR000445-06. Support for this work included funding from the Department of Defense in the Center for Neuroscience and Regenerative Medicine. This project was supported in part by CTSA award No. UL1TR000445 from the National Center for Advancing Translational Sciences. The content is solely the responsibility of the authors and does not necessarily represent the official views of the NCATS, NIH, or NSF.

**Abbreviations**

EMR = Electronic Medical Record; ICD = International Classification of Diseases; PheWAS = Phenotype-Wide Association Study; CPT = Current Procedural Terminology

**Appendix**

Appendix A: List of phrases used for text search to filter healthy radiology reports for MRI studies.

- No focal parenchymal lesions are seen
- No acute intracranial abnormalities.
- No restricted diffusion
- Normal non-contrasted MRI brain
- No acute intracranial abnormality
- Normal brain MRI

- Negative brain MRI
- Normal brain parenchyma
- Normal parenchyma
- Normal MRI of the brain
- Negative MR brain
- No focal lesions or acute process is seen
- No finding of intracranial pathology
- Normal brain MR exam
- Negative MRI brain without contrast
- otherwise negative MRI brain
- Normal MR appearance of the brain
- Normal contrasted MRI of the brain
- Negative MRI brain with and without contrast

Appendix B: List of phrases used for text search to filter healthy radiology reports for CT studies.

- No findings of acute intracranial abnormality
- No findings of intracranial abnormality
- Normal unenhanced head CT.
- No evidence of acute intracranial injury
- No acute process is seen on this unenhanced scan of the brain.
- Normal brain CT
- Normal head CT
- Negative noncontrasted CT of the head.
- Normal head CT examination.
- No visible intracranial injury or other abnormality.
- Normal unenhanced CT of the head
- Normal CT of the brain
- Normal CT
- Normal head CT scan without contrast